\documentclass[%
 reprint,
 amsmath,amssymb,
 aps,
 superscriptaddress,
 pre,
 twocolumn,
 titling,
]{revtex4}

\usepackage{graphicx}
\usepackage{dcolumn}
\usepackage{bm}
\usepackage{braket}

\newcommand{\nc}{\newcommand}
\nc{\p}[2]{\frac{\partial #1}{\partial #2}}

\begin{document}

\preprint{APS/123-QED}
\title{Optimal finite-time Brownian Carnot engine}

\author{Adam G. Frim}%
\email{adamfrim@berkeley.edu}
\affiliation{%
 Department of Physics, University of California, Berkeley, Berkeley, California, 94720
}%
\author{Michael R. DeWeese}
\email{deweese@berkeley.edu}
\affiliation{%
 Department of Physics, University of California, Berkeley, Berkeley, California, 94720
}%
\affiliation{%
Redwood Center For Theoretical Neuroscience,  University of California, Berkeley, Berkeley, California, 94720
}
\affiliation{%
Helen Wills Neuroscience Institute,University of California, Berkeley, Berkeley, California, 94720
}%

\date{\today}

\begin{abstract}
Recent advances in experimental control of colloidal systems have spurred a revolution in the production of mesoscale thermodynamic devices. Functional ``textbook" engines, such as the Stirling and Carnot cycles, have been produced in colloidal systems where they operate far from equilibrium. Simultaneously, significant theoretical advances have been made in the design and analysis of such devices. Here, we use methods from thermodynamic geometry to characterize the optimal finite-time, nonequilibrium cyclic operation of the parametric harmonic oscillator in contact with a time-varying heat bath, with particular focus on the Brownian Carnot cycle. 
We derive the optimally parametrized Carnot cycle, along with two other new cycles and compare their dissipated energy, efficiency, and steady-state power production against each other and a previously tested experimental protocol for the Carnot cycle. We demonstrate a 20\% improvement in dissipated energy over previous experimentally tested protocols and a $\sim$50\% improvement under other conditions for one of our engines,
while our final engine is more efficient and powerful than the others we considered. Our results provide the means for experimentally realizing optimal mesoscale heat engines. 
\end{abstract}

\maketitle


\textit{Introduction}. Since the turn of the millennium, our understanding of nonequilibrium processes has improved dramatically \cite{1997_PRL_Jarzynski,Sekimoto,1999_PRE_Crooks,2005_PRL_Seifert,2010_PRL_Esposito_three,2010_PRL_Sagawa,2012_RPP_Seifert}. Over the past decade in particular, powerful techniques for controlling colloidal mesoscopic systems have facilitated experimental realizations of finite-time thermodynamic cycles~\cite{2006_PRL_Blickle,2011_NatPhys_Blickle,2016_NatPhys_Martinez, 2014_NC_Quinto, 2016_NPhys_Krishnamurthy, 2017_SM_Martinez}. A major step forward was achieved by the construction of a mesoscopic Stirling cycle with a harmonically trapped particle suspended in a temperature-controlled, laser heated fluid~\cite{2011_NatPhys_Blickle}. Another significant advance was achieved by placing an electrically charged colloidal particle in an electrostatic field with tunable noise in order to mimic a thermal bath with a continuously varying temperature~\cite{2016_NatPhys_Martinez}. By following alternating adiabatic (constant Shannon entropy) and isothermal strokes, these authors produced a mesoscopic colloidal Carnot cycle. \\
\indent 
However, although a large class of control protocols, including those used in this experiment, reproduce the standard Carnot cycle when performed quasistatically, the choice of a specific temporal parametrization significantly impacts the thermodynamic performance of the engine when operated in finite time. In this Letter, we connect the versatile thermodynamic geometry approach with the colloidal harmonic oscillator used previously~\cite{2011_NatPhys_Blickle,2016_NatPhys_Martinez} and calculate explicit optimal protocols for this important model system. We find improvements in the efficiency, output power, and dissipated energy in steady-state operation for a wide variety of cycle durations. Our results may find use in practical development of 
mesoscale engines. \\
\indent 
\textit{Thermodynamic geometry.}
Thermodynamic length was originally introduced as a notion of metric distance between equilibrium states of a physical system \cite{1975_JCP_Weinhold, 1979_PRA_Ruppeiner, 1983_Salamon_PRL, 1984_JCP_Salamon, 1985_ZPB_Schlogl, 2007_PRL_Crooks}. Since their introduction, thermodynamic length and similar geometric approaches have found wide-ranging applications in various contexts~\cite{1995_RMP_Ruppeiner}, with recent success in optimal control of nonequilibrium systems~\cite{2012_PRL_Sivak, 2012_PRE_zulkowski,2013_PRE_Deffner, 2013_PLOS_Zulkowski,2015_PRE_zulkowski_over,2015_PRE_Zulkowski_quantum,2015_PRE_Rotskoff,2016_JSM_Mandal,2017_PRE_Rotskoff,PRL_2020_Abiuso,2020_entropy_Abiuso,2021_arxiv_watanabe}. We will follow the treatment introduced in a recent pioneering study~\cite{2020_PRL_Brandner} that successfully applied a geometric approach to closed thermodynamic cycles. In particular, we consider a thermodynamic system with two control parameters, $\mathbf{\Lambda} = (T,\lambda)$, operated cyclically, where $T$ is the time-varying temperature of an external bath and $\lambda$ represents a mechanically varied parameter. The work, $W$, and effective energy intake from the heat source~\cite{2020_PRL_Brandner}, $U$, are given
\begin{equation}
\label{eq:work}
W =- \oint dt \Braket{\p{H_\lambda}{\lambda_t} }\dot{\lambda}_t, \ \ \ \ U = \oint dt \braket{ \log \rho_t} \dot{T}_t ,
\end{equation}
where $H_\lambda$ is the system Hamiltonian for a given value of $\lambda$, $\rho$ is the phase space density, $t$ subscripts denote values at a given time $t$, and brackets denote ensemble averages, i.e. phase space integrals against $\rho$ for a given $t$. 

This expression for the work is fairly standard, but the quantity $U$ above may be less familiar ($U$ \emph{does not} represent the internal energy of the system). For any quasistatic, reversible process, a change in the entropy of a system must coincide with an exchange of heat with an external bath by the amount given by the expression for $U$. For systems driven out of equilibrium, the entropy can increase even in the absence of heat exchange, as is the case during a free expansion, for example. This quantity, which upon performing integration by parts converts to $\oint T dS$ in steady-state driving, where $S$ is the system entropy, then includes contributions at each moment from both the actual amount of heat exchanged with the environment plus the amount of heat that would result in the extra increase in $S$ during a corresponding reversible process at temperature $T$.

Thus, the energy irreversibly dissipated over the course of a full cycle, $A$, (alternatively known as the dissipated availability or work) can be written
\begin{equation}
\label{eq:dissipated_availability}
A = U-W \geq 0 ,
\end{equation}
where the inequality arises from the second law. The equality is saturated only for quasistatic driving in which case the phase space density assumes a Boltzmann form at all times:
\begin{equation}
\label{eq:boltzmann}
\rho_{\mathbf{\Lambda}} = e^{-\beta (H_\lambda-F_{\mathbf{\Lambda}})},
\end{equation}
where $F_{\mathbf{\Lambda}}$ is the free energy for a given $\lambda$ and $T$ and $\beta \equiv (k_B T)^{-1}$.

Following~\cite{2012_PRL_Sivak,2012_PRE_zulkowski, 2020_PRL_Brandner}, we consider the system to be operating in the slow-driving regime wherein temporal variations of control parameters are assumed slow relative to the relaxation timescale of the system. Standard dynamic linear response then gives the dissipated energy to lowest order in 
$\dot{\bm{\Lambda}}$ 
\begin{equation}
\label{eq:dissipated_A_g}
A \approx \oint dt \ \dot{\bm{\Lambda}} \cdot \mathbf{g} \cdot \dot{\bm{\Lambda}},
\end{equation}
where $\mathbf{g}$ is defined as the inverse diffusion tensor, given by equilibrium time-correlation functions
\begin{equation}
g_{ij}(t) = \beta(t) \int_0^\infty dt' \braket{\delta X_i(0) \delta X_j(t')}_{\bm{\Lambda}(t')}.
\end{equation}
Here
$X_i(t)$ is the time-dependent thermodynamic force conjugate to control variable $i$: $X_T =- \log \rho_{\mathbf{\Lambda}}$ and $X_\lambda = -\partial H_\lambda/\partial \lambda_t $ and $\delta X = X-\braket{X}$.
The tensor $\mathbf{g}$ is symmetric by construction and can be shown to be positive-semidefinite \cite{2012_PRE_zulkowski,2020_PRL_Brandner} as a consequence of the second law. Because it 
satisfies these conditions, one can interpret $\mathbf{g}$
as a metric tensor, introducing a well-defined notion of geometric distance on cycles in control parameter space. Non-cyclical paths can yield a negative dissipated energy, suggesting 
that entropy production may be a more appropriate quantity to study when considering such processes~\cite{2012_PRE_zulkowski, 2021_PRE_Large}. For an arbitrary closed path $\bm{\phi}$ through control space, we may define the corresponding thermodynamic length $\mathcal{L}_{\bm{\phi}}$ as 
\begin{equation}
\label{eq:thermo_L}
\mathcal{L}_{\bm{\phi}} =\int \sqrt{ d\bm{\phi} \cdot \mathbf{g} \cdot d\bm{\phi}},
\end{equation}
which is independent of parametrization. Beyond the thermodynamic length, a different geometric quantity, the thermodynamic divergence, may be understood as the thermodynamic cost in dissipated energy of a physical operation. The divergence is defined as
\begin{equation}
\label{eq:thermo_D}
\mathcal{D}_{\bm{\phi}} =\tau \int_0^\tau dt  \  \dot{\bm{\phi}} \cdot \mathbf{g} \cdot \dot{\bm{\phi}},
\end{equation}
where now $\bm{\phi}$ is explicitly parametrized by a time-varying protocol with $t \in [0,\tau]$ and the divergence depends on this parametrization. By comparing Eqs. \eqref{eq:dissipated_A_g} and \eqref{eq:thermo_D}, we see that the thermodynamic divergence precisely matches the dissipated energy of a protocol scaled by the protocol duration. Paths and parametrizations that minimize the thermodynamic divergence are therefore minimally dissipative and thermodynamically optimal in that sense. Between any two points, such paths are known as geodesics. Moreover, for any given path in control space that is not a geodesic, there still exists an optimal parametrization that minimizes the divergence, and therefore the dissipated energy, for a fixed protocol duration. Explicitly, comparing Eqs. \eqref{eq:thermo_L} and \eqref{eq:thermo_D}, the Cauchy-Schwarz inequality implies $A = \mathcal{D}/\tau \geq \mathcal{L}^2/\tau$. This bound is saturated only for optimal driving protocols where control parameters are changed in such a way that the quantity $\dot{\bm{\phi}}\cdot\mathbf{g}\cdot \dot{\bm{\phi}}$, which we identity as the instantaneous dissipated power, $\mathcal{P}_\text{diss}$, is constant over the full protocol duration~\cite{2012_PRL_Sivak}.\\
\indent 
\textit{Brownian working substance.}
The parametric harmonic oscillator is often used as the paradigmatic model of colloidal thermodynamic systems and has been successfully applied as the working substance in physical realizations of mesoscopic heat engines~\cite{2011_NatPhys_Blickle,2016_NatPhys_Martinez}. 
This model system consists of a particle of mass $m$ in a harmonic trap with time dependent stiffness  $k(t)$ in contact with a heat bath at temperature $T(t)$, evolving under Langevin dynamics
\begin{equation}
    \label{eq:Langevin}
    m\ddot{z} = -\zeta \dot{z}-k(t)z+\eta(t),
\end{equation}
where $z$ is the position of the particle, $\zeta$ is the friction coefficient, and $\eta(t)$ is Gaussian white noise satisfying
\begin{equation}
    \braket{\eta(t)\eta(t')} = 2\zeta k_BT \delta(t-t'),
\end{equation}
ensuring that the dynamics satisfy detailed balance. With these two control variables, the thermodynamic forces can be expressed as
\begin{equation}
\mathbf{X} = \begin{pmatrix} \frac{1}{2T} (p^2/m+k z^2) +\frac{1}{T} \log \left(2\pi \sqrt{\frac{mT^2}{k}}\right) , & -1/2 z^2\end{pmatrix},
\end{equation}
where $p$ is the momentum of the particle. Following methods similar to \cite{2012_PRE_zulkowski}, we arrive at our first 
major
result, the full metric tensor for this thermodynamic space (see \cite{SM} for a detailed derivation):
\begin{multline}
\label{eq:tensor}
g_{ij} = \beta \int_0^\infty dt \braket{\delta X_i (t)\delta X_j(0)} \\
 = \frac{mk_B}{4\zeta} \begin{pmatrix}
\frac{1}{T} \left(4+\frac{\zeta^2}{km}\right) &
-\frac{1}{ k } \left(2+\frac{\zeta^2}{km}\right) \\
-\frac{1}{ k } \left(2+\frac{\zeta^2}{km}\right) &
\frac{T}{k^2} \left(1+\frac{\zeta^2}{km}\right)
\end{pmatrix}_{ij}.
\end{multline}

\begin{figure*}[t]
\centering{
\includegraphics[width=\textwidth]{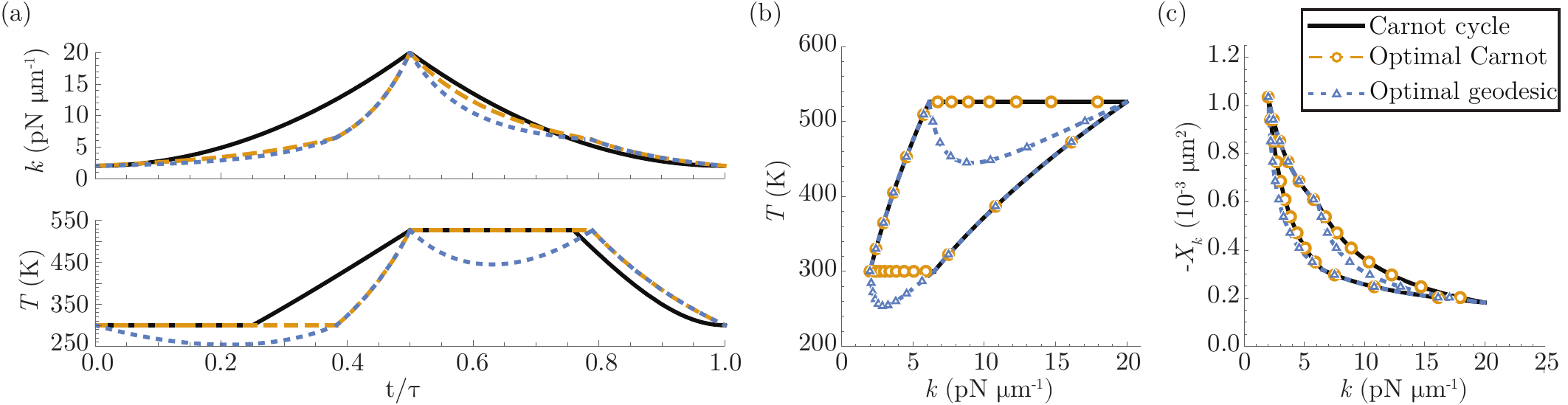}
 \\}
 \caption{Shape and temporal parametrization of a previously tested Brownian Carnot cycle, the optimized Brownian Carnot cycle, and an optimal geodesic cycle matched to the Carnot cycle. \textbf{(a)} The stiffness (top) and temperature (bottom) for three heat engines
 as a function of time: an experimentally tested Brownian Carnot cycle~\cite{2016_NatPhys_Martinez} (solid black), 
 the optimally parametrized Brownian Carnot cycle (dashed orange), and the optimal cycle employing geodesics between adjacent pairs of corners of the previous Carnot cycles (dotted blue). Horizontal axes are shared between subplots. \textbf{(b)} The functional form in control parameter space of the Carnot cycle (solid black) and the optimal geodesic cycle (dotted blue). Orange circles (blue triangles) denote points along the optimal Carnot (geodesic) cycle parametrization at 25 equal duration intervals in time. \textbf{(c)} Same as in \textbf{(b)} for the Clausius diagram of stiffness $k$ against its conjugate thermodynamic force in the quasistatic limit $-X_k = x^2/2 = k_BT/2k$. 
 }
\label{fig:shape}
\end{figure*}

\indent 
\textit{Optimal Brownian Carnot engine.}
The Carnot engine is a four-stroke engine consisting of alternating isothermal steps in contact with a heat bath of either a hot temperature $T_h$ or a cold temperature $T_c<T_h$ and adiabatic steps during which no heat is exchanged with 
a bath. Consistent with the second law, all engines acting reversibly between two heat baths at these temperatures cannot have a lower thermodynamic efficiency than the Carnot engine, with Carnot thermodynamic efficiency given by $\eta_\text{C} = 1-T_c/T_h$. However, the classical Carnot engine operates quasistatically and therefore performs only a finite amount of work over an arbitrarily long cycle period, thus delivering no power. Accordingly, finite-time, nonzero-power thermodynamics of this engine have been a topic of significant interest~\cite{1975_AJP_Curzon,2005_PRL_VandenBroeck,2007_EPL_Schmiedl,2009_PRL_Esposito,2010_PRL_Esposito,2011_PRL_Benenti,2015_PRE_Brandner,2015_PRX_Brandner,2016_PRL_Proesmans,2016_PRL_Shiraishi,2018_PRL_Pietzonka, 2018_PRE_Ma_1,2018_PRE_Ma_2,PRL_2020_Abiuso,2020_PRL_Ma,2021_PRL_Miller}.\\
\indent
Recently, a laser trapped colloidal particle in contact with a time-varying (effective) heat bath was used to experimentally produce a Carnot engine with the Brownian working substance described in the previous section.
Isothermal expansion and compression arise from changes in $k(t)$ for a fixed temperature $T$. Adiabatic paths were achieved by holding fixed $T^2/k$, an adiabatic invariant for this system, which maintains a constant value of Shannon entropy~\cite{2015_PRL_Martinez}.
A portrait of the Carnot cycle in $(T,k)$ space is plotted in Fig.~\ref{fig:shape}(b). \\
\indent 
Considering the metric given in Eq.~\eqref{eq:tensor}, we now construct the optimal parametrization for the Carnot engine. For isothermal steps, the dissipated power can be obtained by setting $\dot{T} = 0$ in the integrand of Eq.~(\ref{eq:dissipated_A_g}), which gives
\begin{equation}
\label{eq:isothermal_power}
\mathcal{P}_{\text{diss,iso}} = \frac{mk_BT}{4\zeta} \frac{\dot{k}^2}{k^2} \left(1+\frac{\zeta^2}{km}\right).
\end{equation}
The optimal trajectory in $k$ is then found through the corresponding Euler-Lagrange equation
\begin{equation}
\label{eq:isothermal_EL}
\left(2\zeta+\frac{2m k}{\zeta}\right) \ddot{k} - \left(\frac{3\zeta}{k}+\frac{2m}{\zeta}\right)\dot{k}^2 = 0,
\end{equation}
which can be solved numerically for given initial and final values of $k$. The total thermodynamic cost of such an isothermal step is given by the integral of Eq.~\eqref{eq:isothermal_power} for the solution $k_{\text{iso}}^*$ of Eq.~\eqref{eq:isothermal_EL} over the duration of the protocol. \\
\indent
For adiabatic steps, the key constraint is that $\alpha \equiv T^2/k$ be held fixed. Under this constraint, the dissipated power is given by
\begin{equation}
\label{eq:adiabatic_power}
 \mathcal{P}_{\text{diss}} = \frac{k_B\zeta \alpha \dot{T}^2}{4 T^{3}},
\end{equation}
leading to the Euler-Lagrange equation,
\begin{equation}
\label{eq:adiabatic_EL}
\ddot{T} = \frac{3\dot{T}^2}{2T},
\end{equation}
which is analytically solvable. For a generic adiabatic protocol of duration $\tau$ that transitions from initial stiffness $k_i$ and temperature $T_i \equiv \sqrt{\alpha k_i}$ at time $t=0$ to a final stiffness $k_f$ and temperature $T_f \equiv \sqrt{\alpha k_f}$ at time $t=\tau$, the protocol is 
\begin{align}
\label{eq:adiabatic_protocol}
&T_{\text{adiab}}^*(t) = \frac{T_iT_f \tau^2}{(\sqrt{T_f}\tau + (\sqrt{T_i} -\sqrt{T_f})t)^2},\\
&k_{\text{adiab}}^*(t) = T_{\text{adiab}}^*(t)^2 /\alpha,
\end{align}
leading to a (constant) energetic cost of
\begin{equation}
\label{eq:adiabatic_power_optimal}
 \mathcal{P}_{\text{diss,adiab}}^* =\frac{ k_B\zeta \alpha (\sqrt{T_f}-\sqrt{T_i})^2}{T_iT_f \tau^2}.
\end{equation}
 \\
\indent 
The total Carnot cycle consists of alternating isothermal and adiabatic processes connecting four points in $(T,k)$ space. Following~\cite{2016_NatPhys_Martinez}, we order the four strokes as: 1) isothermal compression at $T_c$, 2) adiabatic compression between $T_c$ and $T_h$, 3) isothermal expansion at $T_h$, and 4) adiabatic expansion between $T_h$ and $T_c$. In order to allocate the optimal amount of time to each stroke of the cycle, we note that for any optimally-parametrized process $\bm{\phi}^*$, the dissipated energy is given $A_{\bm{\phi}^*}= \mathcal{D}_{\bm{\phi}^*}/\tau =\mathcal{L}_{\bm{\phi}^*}^2/\tau$, such that $\mathcal{P}_{\text{diss},\bm{\phi}^*} = \mathcal{L}_{\bm{\phi}^*}^2/\tau^2$. This must be true of both the full cycle and each optimized stroke, such that $\mathcal{L}_i^2/\tau_i^2 = \mathcal{L}_\text{tot}^2/\tau^2 \implies \mathcal{L}_i/\tau_i = \mathcal{L}_\text{tot}/\tau \implies \tau_i = \tau \mathcal{L}_i/\mathcal{L}_{\text{tot}}$ where $\tau_i$ and $\mathcal{L}_i$ are the duration and thermodynamic length, respectively, of the $i$th stroke ($\sum_i \tau_i = \tau$).\\
\begin{figure*}[t!]
{\centering\includegraphics[width=\textwidth]{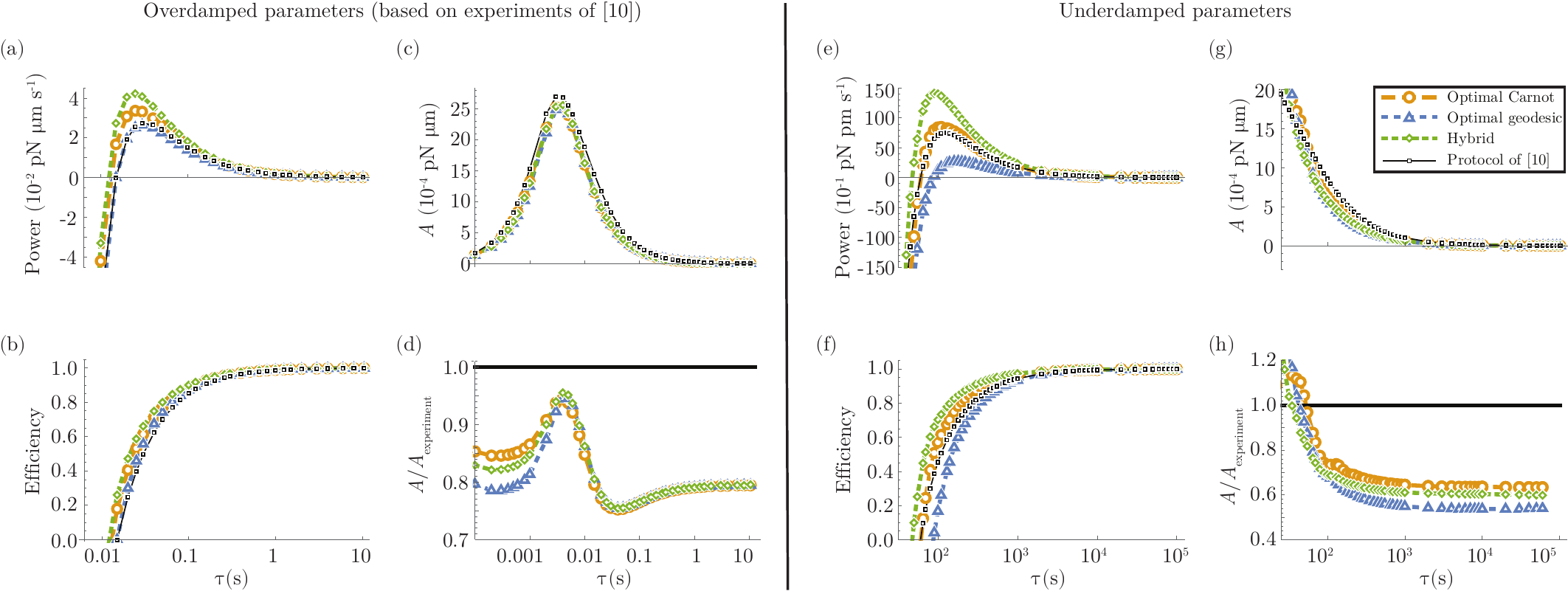}}
\caption{Performance of various engines considered in the main text as a function of protocol duration $\tau$. We numerically simulate the system under cyclic operation of the control parameters for protocol duration $\tau$ and plot the performance of each engine after reaching steady state. In all plots, results of simulations of an engine following the experimental protocol used in \cite{2016_NatPhys_Martinez} are shown with black squares and a straight black linear interpolation, the optimal Carnot cycle with orange circles and dashed orange linear interpolation, the optimal geodesic cycle with blue triangles and blue dotted linear interpolation, and a hybrid cycle consisting of the optimized Carnot cycle with the cold isothermal compression step replaced by the corresponding geodesic path with green diamonds and dashed linear interpolation. \textbf{(a)} Average power. \textbf{(b)} Average efficiency. \textbf{(c)} Average dissipated energy. \textbf{(d)} Dissipated energy scaled relative to that of experimental protocol. Material parameters, i.e. $m, \zeta, T_c, T_h$ and $k$, are chosen based on those used in the experiment of \cite{2016_NatPhys_Martinez}. Observe the dissipation is minimized for the geodesic engine, though this engine is less efficient and less powerful than the optimized Carnot and hybrid engines; the hybrid engine is both most efficient and powerful for all protocol durations considered. \textbf{(e)}-\textbf{(h)} Same as \textbf{(a)}-\textbf{(d)} for different material parameters for which the contrasting performance between different cycles is clearer (see \cite{SM} for details of material parameter values). Horizontal axes of \textbf{(a)},\textbf{(c)},\textbf{(e)}, and \textbf{(g)} are the same as each subfigure below them, \textbf{(b)},\textbf{(d)},\textbf{(f)}, and \textbf{(h)}, respectively.}
\label{fig:performance}
\end{figure*}
\textit{Optimal geodesic engine.}
The Carnot engine is significant in reversible thermodynamics as the paradigmatic model of a heat engine with maximal efficiency under various conditions, such as for specified hot and cold heat baths and considering only inward heat flows when assessing the thermodynamic cost. As discussed previously, its path in control space follows a prescribed shape set by the well-known quasistatic cycle. However, in this geometric framework, there is (at least) one well-defined minimal dissipation path connecting any two points in control space: a geodesic. Given the existence of geodesic protocols that are less dissipative than the corresponding strokes of the Carnot cycle, we next construct an engine consisting only of geodesics that connect adjacent pairs of the four corners of the Carnot cycle. The shape of such an engine is depicted in Fig.~\ref{fig:shape}(b). The two adiabatic strokes are in fact exactly geodesic and are unchanged for this cycle, though the geodesic paths corresponding to the isothermal strokes now involve variations in temperature. Individual strokes again may be pieced together according each stroke a duration proportional to its thermodynamic length as required for optimally-parametrized processes, leading to the parametrization of the cycle depicted in Fig. \ref{fig:shape}(a). \\
\indent \textit{Engine performance.} We now calculate the performance of the various engines described in the previous section and compare them against a previously studied experimental cycle~\cite{2016_NatPhys_Martinez}, which we use as a benchmark. To mimic the experiment, we pin the four corners of the Carnot cycle at $(T_0,k_0) = (300 \text{ K}, 2 \text{ pN $\mu$m${}^{-1}$)}$, $ (T_1,k_1) = (300 \text{ K} , 6.5 \text{ pN $\mu$m${}^{-1}$)}$, $(T_2,k_2) = (600\sqrt{10/13} \text{ K}, 20  \text{ pN $\mu$m${}^{-1}$)}$, $(T_3,k_3)=  (600\sqrt{10/13} \text{ K}, 80/13  \text{ pN $\mu$m${}^{-1}$)}$,  and $(T(\tau),k(\tau)) =(T_0, k_0)$. Following standard methods starting from Eq.~\eqref{eq:Langevin}, we can derive an equivalent Fokker-Planck equation for the evolution of the probability density over phase space. Integrating across various covariances~\cite{SM}, we arrive at the coupled differential equations governing the evolution of $\braket{z^2}$, $\braket{pz}$, and $\braket{p^2}$:
\begin{align}
&\frac{d}{dt} \braket{z^2} = \frac{2}{m} \braket{pz},\\
&\frac{d}{dt} \braket{pz} = \frac{\braket{p}^2}{m} - k\braket{z^2} - \frac{\zeta}{m} \braket{pz},\\
&\frac{d}{dt} \braket{p^2} = 2k\braket{pz}-2\frac{\zeta}{m}\braket{p^2} + 2\zeta k_BT.
\end{align}
Given that our model system only encounters a harmonic potential, if we assume the system starts in a Gaussian form, it will remain Gaussian for the duration of the protocol, such that these covariances encode the \textit{entire} phase space distribution of the Brownian oscillator. Therefore, numerically solving these equations for given control parameters $k(t)$ and $T(t)$ and allowing the system to come to its steady state, we are able to fully simulate the system and evaluate various performance metrics of the engine. We use \textit{simulations} of the experimental protocol of~\cite{2016_NatPhys_Martinez} as a benchmark. We do this rather than use the actual experimental results to allow for the evaluation of a greater range of protocol durations and for more detailed (simulated) data than was experimentally measured.
We validate our numerical simulations by direct comparison to the experiment in the Supplemental Materials \cite{SM}.

In Fig.~\ref{fig:performance}(a), we plot the power output $P=W/\tau$ as a function of cycle duration $\tau$. As expected, for small times $W$ is negative such that no work is extracted from the cycle and the power is large in magnitude and negative for 
each of the engines, resulting in a maximum positive value for the power at a finite value of cycle duration. We observe a noticeable benefit to the use of the optimized Carnot engine, though the optimal geodesic engine actually leads to a \textit{reduction} in power. 

To better understand this, we plot the dissipated energy $A$ for these protocols in Fig.~\ref{fig:performance}(c)-(d). We note a $>$$20$$\%$ decrease in the dissipated work per cycle for both the optimal Carnot engine as well as the geodesic engine compared to the previously tested experimental protocol. For these experimental parameters, the thermodynamic length of the optimal Carnot cycle is only $0.0003\%$ greater than that of the geodesic cycle, such that its benefit is difficult to observe in these figures. We therefore also simulate for a different set of material values where the benefits are clearer, now demonstrating a $\sim$50\% decrease in dissipated work for the optimal geodesic engine; these results are plotted in Fig. \ref{fig:performance}(e)-(h). This second set of material parameters corresponds to using a ball of millimeter radius and density comparable to gold, which may not be achievable with current experimental technology but serves to illustrate the differences between the various cycles in a significantly more underdamped regime. See \cite{SM} for further simulation details. We can compute the efficiency of all of these engines, displayed in Fig. \ref{fig:performance}(b) and (f) as a function of cycle duration. Here we define efficiency not as the ratio of work output to heat input, but as the ratio of work output to total effective thermal energy uptake~\cite{2020_PRL_Brandner}:
\begin{equation}
\label{eq:efficiency}
\eta = \frac{W}{U} \approx 1-\frac{A}{\mathcal{W}},
\end{equation}
where $\mathcal{W}$ is the net work extracted from a quasistatic cycle with $\rho_t = \rho_{\mathbf{\Lambda}}$ at all time; Eq.~(\ref{eq:efficiency}) holds to lowest nontrivial order in driving rates. As stated previously, this definition is more appropriate for engines of time-variable heat baths and has a universal maximum value of one for any reversible engine. As expected, we observe that both optimized protocols yield a superior efficiency relative to the experimental protocol in the long time limit, as well as for earlier times in most cases. Interestingly, the optimal Carnot cycle is more efficient than the optimal geodesic cycle. This surprising result, along with the comparable result found for the powers, may be understood by recognizing that the cost function for this optimization scheme is only the availability $A$, and indeed we see the geodesic cycle produces the smallest total availability. In fact, for slowly driven cycles, the optimal geodesic cycle yields the minimum possible value of $A$ for any cycle passing through the four corners of the Carnot cycle in a given time. However, the efficiency and power depend not only on the dissipated energy but also on the total possible energy accessible to the engine in the most ideal conditions, namely $\mathcal{W}$. This therefore introduces a further figure of merit of an engine: for slowly driven systems, the most efficient engine will minimize the ratio $A/\mathcal{W}$. Thus, although the (optimal) Carnot engine produces a larger dissipated energy, it is more efficient. 

\begin{figure}[h]
\centering{
\includegraphics[width=\columnwidth]{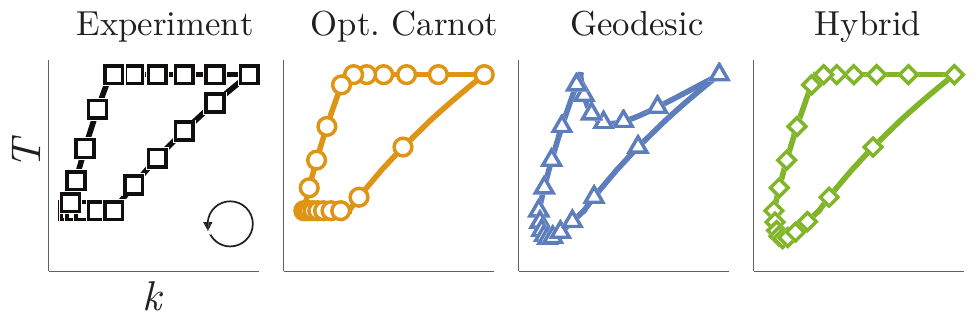}
 \\}
 \caption{The thermodynamic cycles and parametrizations we consider. Closed curves represent the bath temperature ($T$) vs. trap stiffness ($k$) for each engine. Markers are placed at 20 points along each cycle separated by equal time intervals. Colors and marker shapes correspond to the conventions in previous figures.
 }
\label{fig:all_shapes}
\end{figure}

In fact, given that $\mathcal{W}$ is a monotonic function of the area enclosed by the cycle plotted in Fig.~\ref{fig:shape}(b), the geodesic corresponding to the hot isothermal expansion acts to decrease the value of $\mathcal{W}$ as it bows downward into the the cycle, whereas the equivalent for cold isothermal compression acts to increase $\mathcal{W}$. This intuition suggests that we form a more efficient cycle than any considered above by starting with the optimized Carnot cycle and replacing only the cold isothermal compression with the corresponding geodesic path. We plot the shape of this cycle in comparison to others we consider in Fig.~\ref{fig:all_shapes}.
Indeed, of all the cycles we considered, this hybrid cycle produces the most efficient engine in terms of both $A/\mathcal{W}$ and power (Fig.~\ref{fig:performance}). See \cite{SM} for further comparisons of all simulated cycles to the experimental results of \cite{2016_NatPhys_Martinez}. All of the optimized cycles statistically significantly outperform the experimentally measured Carnot cycle.

\textit{Discussion.}
Here we have considered the optimal temporal parametrization for the Brownian Carnot cycle, and we have also  derived two new finite-time thermodynamic cycles that incorporate geodesics connecting consecutive pairs of corners of the Carnot cycle.
We have demonstrated that each of these three new engines is less-dissipative than the experimentally tested temporal parametrization of the Brownian Carnot cycle \cite{2016_NatPhys_Martinez}, with our hybrid engine being the most efficient among those studied. A clear next step would be to carry out a similar procedure for other thermodynamic engines and refrigerators, such as those corresponding to the Stirling and Otto cycles.

Although the treatment of thermodynamic length is highly accurate for slowly driven cycles \cite{SM}, contributions from higher order corrections have been reported in the literature \cite{wadia} and have been used, for example, to give stronger bounds on free energy 
estimates of thermodynamic processes~\cite{2020_JCP_Blaber}. Likewise, lessons from studies of finite-time processes that shortcut relaxation timescales could further facilitate the development of optimal cyclic engines \cite{2007_PRL_Schmiedl,2016_NP_Martinez_ESE,2017_PRE_Jarzynski,2017_PRE_Geng,2017_NJP_Patra,2018_NJP_Chupeau,2019_RMP_Guery,2020_SciP_Salambo, 2020_PRE_Baldassarri, 2021_NP_Iram,2021_PRE_Frim}; such questions have been studied for overdamped dynamics \cite{PRE_2020_Plata,JSM_2020_Plata}

Finally, although we were able to construct a novel and minimally-dissipative cycle connecting each of the four pairs of adjacent corners of the Carnot cycle with geodesics, it was \textit{less} efficient than the corresponding Carnot cycle. 
Recognizing that we could simultaneously reduce the availability $A$ and increase the work $W$ by incorporating a geodesic path in place of the cold isothermal compression while retaining the other three strokes of the optimized Carnot cycle, we obtained our hybrid cycle, which is the most efficient of the engines we considered. We found this more efficient cycle by assembling strokes that were obtained by minimizing only dissipation, but it may be possible to derive maximally efficient cycles more directly. Importantly, for slow driving, the form of efficiency we considered here (Eq.~\ref{eq:efficiency}) is a purely geometric quantity, such that its optimization suggests a new design principle for the \textit{shape} of a cycle, beyond just its optimal temporal parametrization or introducing geodesics between pairs of points. We will pursue this further in future work.

\textit{Conclusion.} In this Letter, we have characterized the thermodynamic geometry of the colloidal harmonic oscillator system and used these results to derive explicit protocols for optimal Carnot-like cycles. Using similar methods, one could construct optimal parametrizations and introduce geodesics to minimize dissipation for any chosen cycle of the model working system studied. Future work may be directed towards higher order corrections to our results, application of our results to the study of further cycles, and the development of fundamentally new and more efficient nonequilibrium thermodynamic cycles. We hope our results will facilitate the design and construction of optimal thermal machines at the mesoscale.

\begin{acknowledgements}
The authors would like to thank Adrianne Zhong and David Sivak for useful discussions, Mart\'{i} Perarnau-Llobet for helpful comments on the manuscript, and Ra\'{u}l Rica for granting us access to and assistance in analyzing the data collected in \cite{2016_NatPhys_Martinez}. AGF is supported by the NSF GRFP under Grant No. DGE 1752814. This work was supported in part by the U. S. Army Research Laboratory and the U. S. Army Research Office under contract W911NF-20-1-0151.
\end{acknowledgements}

\end{document}


\preprint{APS/123-QED}
\title{Supplementary Material for ``Optimal finite-time Brownian Carnot engine"}

\author{Adam G. Frim}%
\affiliation{%
 Department of Physics, University of California, Berkeley, Berkeley, California, 94720
}%
\author{Michael R. DeWeese}
\affiliation{%
 Department of Physics, University of California, Berkeley, Berkeley, California, 94720
}%
\affiliation{%
Redwood Center For Theoretical Neuroscience,  University of California, Berkeley, Berkeley, California, 94720
}
\affiliation{%
Helen Wills Neuroscience Institute,University of California, Berkeley, Berkeley, California, 94720
}%

\date{\today}

\maketitle

Throughout the Supplementary Material, all equations and figures are listed are notated with an ``S" before the appropriate number. Any reference to equations or figures without an S refers to one in the main text.

\section{Thermodynamic metric for the parametric harmonic oscillator}
Here, we will derive the thermodynamic metric given in Eq.~(11) of the main text. Significant methods in our treatment were previously developed in \cite{2012_PRL_Sivak, 2012_PRE_zulkowski}, and we will follow a similar derivation. First, we recapitulate the definition 
\begin{equation}
    \label{eq:metric_def}
    g_{ij} = \beta \int_0^\infty \braket{\delta X_i(t) \delta X_j(0)},
\end{equation}
where $\delta X_i  \equiv X_i-\braket{X_i }$, $\beta \equiv (k_BT)^{-1}$ is inverse temperature, and brackets denote an ensemble average over an initial equilibrium distribution and thermal noise. Following Eq.~(10) of the main text, these thermodynamic forces take the form
\begin{equation}
\mathbf{X} = \begin{pmatrix} \frac{1}{2T} (p^2/m+k z^2) +\frac{1}{T} \log \left(2\pi \sqrt{\frac{mT^2}{k}}\right) , & -1/2 z^2\end{pmatrix}.
\end{equation}
where $z$ is the particle's position, $m$ is mass, $k$ is the stiffness constant, and $T$ is the temperature of the heat bath. In order to calculate the metric, therefore, we need to calculate various two-point correlation functions, which are dynamical quantities for fixed external parameters. In particular, we will be interested in $z$ and $p\equiv m\dot{z}(t)$ for a particle whose dynamics are given by Eq.~(8) of the main text at fixed values of $k$ and $T$, reproduced here 
\begin{equation}
    \label{eq:dynamics}
    m\ddot{z} +kz+\zeta \dot{z} = F(t).
\end{equation}
Given that this is a second order linear differential equation, we may write its solution as the sum of homogeneous, $z^{(h)}$, and particular, $z^{(p)}$, solutions, $z = z^{(h)}+z^{(p)}$, where all dependence on initial conditions is carried in $z^{(h)}$ and all dependence on the random thermal forcing $F(t)$ is carried in $z^{(p)}$. Following this reasoning and assuming $F(t)$ is a zero-mean Gaussian white noise process, it is a simple exercise to show that in general $\braket{\delta X_i(t) \delta X_j(0)} = \braket{\delta X_i^{(h)}(t) \delta X_j(0)}$ for any thermodynamic force $X$. Therefore, we need only calculate the homogeneous solution to the differential equation, 
and thus set $F=0$.

First, let us consider the case 
$\zeta^2 \neq 4km$, such that Eq.~\eqref{eq:dynamics} with $F=0$ is easily solved to give  
\begin{equation}
z_h(t)=
    \label{eq:z_not_critical}
    \frac{p(0)+m\Lambda_-z(0)}{m(\Lambda_--\Lambda_+)}e^{-\Lambda_+t}+\frac{p(0)+m\Lambda_+z(0)}{m(\Lambda_+-\Lambda_-)}e^{-\Lambda_-t}
\end{equation}
for given initial conditions $z(0)$ and $p(0)$ and 
\begin{equation}
\Gamma_\pm \equiv \frac{\zeta}{2m}\pm \frac{1}{2} \sqrt{\left(\frac{\zeta}{2}\right)^2-\frac{4k}{m}}.
\end{equation}
$p(t)$ is then given by $m\dot{z}$. Assuming the system is initially Boltzmann distributed, $\rho_{\text{Boltzmann}}(z,p) \propto \text{exp}[-\beta(p^2/2m+kz^2/2)] $, equilibrium correlation functions 
can then be written as follows:
\begin{subequations}
\begin{align}
    &\braket{z(0)^2} = \frac{k_BT}{k}\\
    &\braket{p(0)^2} =  mk_BT \\
    &\braket{z(0)^4} =  3\left(\frac{k_BT}{k}\right)^2\\
    &\braket{z(0)^2p(0)^2} = \braket{z(0)^2}\braket{p(0)^2} = (k_BT)^2 \frac{m}{k}\\
    &\braket{p(0)^4} =  3\left( mk_BT \right)^2,
\end{align}
\end{subequations}
and we find the following averages
\begin{subequations}
\begin{align}
    &\braket{\delta z_h(t)^2 \delta z(0)^2} = \braket{(z_h(t)^2-\braket{z(0)^2})(z(0)^2-\braket{z(0)^2})} = 2 \left(\frac{k_BT}{k(\Lambda_+-\Lambda_-)}\right)^2 (\Lambda_-e^{-\Lambda_+t}-\Lambda_+e^{-\Lambda_-t})^2\\
    &\braket{\delta p_h(t)^2 \delta z(0)^2} = \braket{(p_h(t)^2-\braket{p(0)^2})(z(0)^2-\braket{z(0)^2})} = 2 \left(\frac{mk_BT\Lambda_+\Lambda_-}{k(\Lambda_+-\Lambda_-)}\right)^2 (e^{-\Lambda_+t}-e^{-\Lambda_-t})^2\\
    &\braket{\delta z_h(t)^2 \delta p(0)^2} = \braket{(z_h(t)^2-\braket{z(0)^2})(p(0)^2-\braket{p(0)^2})} = 2 \left(\frac{mk_BT}{\Lambda_+-\Lambda_-}\right)^2 (e^{-\Lambda_+t}-e^{-\Lambda_-t})^2\\
    &\braket{\delta p_h(t)^2 \delta p(0)^2} = \braket{(p_h(t)^2-\braket{p(0)^2})(p(0)^2-\braket{p(0)^2})} = 2 \left(\frac{mk_BT}{\Lambda_+-\Lambda_-}\right)^2 (\Lambda_-e^{-\Lambda_-t}-\Lambda_+e^{-\Lambda_+t})^2,\\
\end{align}
\end{subequations}
which, upon integration yield
\begin{subequations}
\label{eq:covs_1}
\begin{align}
    &\int_0^\infty dt \braket{\delta z_h(t)^2 \delta z(0)^2} = \frac{m(k_BT)^2}{k^2 \zeta} \left(1+\frac{\zeta^2}{km}\right)\\
    &\int_0^\infty dt \braket{\delta p_h(t)^2 \delta z(0)^2} = \int_0^\infty dt \braket{\delta z_h(t)^2 \delta p(0)^2} =\frac{(mk_BT)^2}{k\zeta} \\
    &\int_0^\infty dt \braket{\delta p_h(t)^2 \delta p(0)^2} =\frac{m^3(k_BT)^2}{\zeta},\\
\end{align}
\end{subequations}
which we can now use to calculate the thermodynamic metric. Before doing so, we must also consider the unique case of critical damping for which $\zeta^2 = 4mk \implies \Lambda_+ = \Lambda_-$ and the dynamics yield an entirely different homogeneous solution:
\begin{equation}
z_{h,\text{crit}}(t)=
    \label{eq:z_critical}
    e^{-\sqrt{\frac{k}{m}}t} \left[ y(0) + \left(z(0)\sqrt{\frac{k}{m}} +\frac{z(0)}{m}\right)t  \right].
\end{equation}
Carrying out the same procedure yields
\begin{subequations}
\begin{align}
    &\braket{\delta z_{h,\text{crit}}(t)^2 \delta z(0)^2} = \frac{2}{m}\left(\frac{k_BT}{k}\right)^2 (m+2\sqrt{km}t +kt^2)e^{-2\sqrt{\frac{k}{m}}t} \\
    &\braket{\delta p_{h,\text{crit}}(t)^2 \delta z(0)^2} =\braket{\delta z_{h,\text{crit}}(t)^2 \delta p(0)^2}= 2(k_BT)^2t^2e^{-2\sqrt{\frac{k}{m}}t} \\
    &\braket{\delta p_{h,\text{crit}}(t)^2 \delta p(0)^2} =2m(k_BT)^2 (m-2\sqrt{km}t +kt^2)e^{-2\sqrt{\frac{k}{m}}t} ,
\end{align}
\end{subequations}
and 
\begin{subequations}
\label{eq:covs_2}
\begin{align}
    &\int_0^\infty dt \braket{\delta z_h(t)^2 \delta z(0)^2} = \frac{5(k_BT)^2 \sqrt{km}}{2k^3} =  \frac{m(k_BT)^2}{k^2 \zeta} \left(1+\frac{\zeta^2}{km}\right)\\
    &\int_0^\infty dt \braket{\delta p_h(t)^2 \delta z(0)^2} =  \int_0^\infty dt \braket{\delta z_h(t)^2 \delta p(0)^2}=\frac{m(k_BT)^2  \sqrt{km}}{2k^2} =\frac{(mk_BT)^2}{k\zeta} \\
    &\int_0^\infty dt \braket{\delta p_h(t)^2 \delta p(0)^2} =\frac{(k_BT)^2 m^2 \sqrt{km}}{2k} = \frac{m^3(k_BT)^2}{\zeta},\\
\end{align}
\end{subequations}
such that, ultimately, regardless of the value of $k$, we arrive at the same time correlation functions. We must calculate correlations between various pairs of the conjugate forces
\[\delta\mathbf{X} =
\begin{pmatrix} 
\frac{1}{2T} (\delta p^2/m+k\delta z^2) 
& -\frac{1}{2}\delta z^2
\end{pmatrix}.
\]
Combining Eqs~\eqref{eq:metric_def},\eqref{eq:covs_1}, and \eqref{eq:covs_2}, we finally have 
\begin{equation}g_{ij} = 
\begin{pmatrix}
    g_{TT} & g_{T\lambda} \\ g_{\lambda T} & g_{\lambda\lambda} 
\end{pmatrix}_{ij}
=\frac{mk_B}{4\zeta} \begin{pmatrix}
\frac{1}{T} \left(4+\frac{\zeta^2}{km}\right) &
-\frac{1}{ k } \left(2+\frac{\zeta^2}{km}\right) \\
-\frac{1}{ k } \left(2+\frac{\zeta^2}{km}\right) &
\frac{T}{k^2} \left(1+\frac{\zeta^2}{km}\right) 
\end{pmatrix}_{ij},
\end{equation}
reproducing Eq. (11) of the main text.

\section{Fokker-Planck dynamics of the harmonic oscillator and simulations}
Here we present a 
derivation of the system of equations used to simulate the harmonic oscillator and we provide some details of those simulations. Following standard methods \cite{Kadanoff}, we may translate the Langevin equation given in Eq.~\eqref{eq:dynamics} into a Fokker-Planck equation over the particle's distribution in phase space:
\begin{equation}
    \label{eq:fp}
    \frac{\partial \rho}{\partial t} + \frac{p}{m} \frac{\partial \rho }{\partial z}-kz \frac{\partial \rho}{\partial p} - \frac{\zeta}{m} \frac{\partial (p\rho)}{\partial p}-\frac{\zeta}{\beta} \frac{\partial^2 \rho}{\partial p^2} = 0.
\end{equation}
Integrating across various dynamical variables leads to evoluation equations for the covariances. Explicitly, let us first integrate against $z^2$. All terms that do not involve derivatives with respect to $z$ trivially die as $\rho$ is normalized and therefore must be zero at the (infinite) boundaries. This then leaves
\[\left(\frac{\partial \rho}{\partial t} + \frac{p}{m} \frac{\partial \rho }{\partial z}\right)z^2 \implies  \frac{d}{dt} \braket{z^2} = -\frac{1}{m}\int dz dp \  pz^2 \frac{\partial \rho }{\partial z} =  \frac{2}{m} \int dz dp pz \rho = \frac{2}{m}\braket{pz},
\]
where we have integrated by parts in the intermediate steps. Similar arguments lead to the rest of the coupled system in Eqs.~(19)-(21) of the main text. 

We reiterate here that, for the dynamics considered, if the system is initially Gaussian distributed, it will remain so for all times, such that knowledge of these covariances specifies the entire phase space distribution:
\[
\rho_{\text{Gaussian}} = \frac{1}{2\pi\sqrt{\braket{p^2}\braket{z^2}-\braket{pz}^2}}
\text{exp}\left[ -\frac{\braket{p^2}z^2 +\braket{z^2}p^2 -2\braket{pz}pz}{2(\braket{p^2}\braket{z^2} - \braket{pz}^2)}\right],
\]
from which one may derive the free energy, entropy, and all other thermodynamic quantities of interest. As a result, with these three covariances alone, we can faithfully study the system well outside of the linear response regime. 

Simulations are then done by numerically integrating Eqs. (19)-(21) for a given set of parameter values. In particular, we first fix $m$ and $\zeta$ and then repeatedly integrate Eqs. (19)-(21), where $k(t)$ and $T(t)$ are cyclically varied with a set time period $\tau$ according to the protocol in question (e.g., according to the optimal Carnot engine or the geodesic engine) until the engine arrives at a steady state. For protocols with very 
small
values of $\tau$, reaching the steady state may take a significant number of cycles, whereas for large values where the cycle is near-equilibrium, steady state cyclic operation can be accomplished after only a handful of cycles. All thermodynamic quantities are then calculated under steady state operation, again by means of numerical integration. All integrations are carried by the NIntegrate function of Wolfram Mathematica 12. See Table~S1 for details of the parameter values used to generate Fig.~2 of the main text. 

\begin{table}[h!]
    \centering
    \[\begin{tabular}{|c||c|c|}
    \hline
        Parameter & Fig. 2(a)-(d) & Fig. 2(e)-(h)\\
        \hline
        \hline
         $m$ & 0.545 pg  & 80.9 mg \\
         $\zeta$ &7.51 $\mu$g  s${}^{-1}$ & 15.0 mg s${}^{-1}$\\
         \hline
         $T_0$ & 300 K &  300 K\\
         $T_1$ & 300 K&  300 K\\
         $T_2$ & 526.235 K &  526.235 K\\
         $T_3$ & 526.235 K &  526.235 K\\
         \hline
         $k_0$ & 2.00 mg s${}^{-2}$& 6.40 mg s${}^{-2}$\\
         $k_1$ & 6.50 mg s${}^{-2}$ & 20.8 mg s${}^{-2}$\\
         $k_2$ & 20.0 mg s${}^{-2}$ & 64.0 mg s${}^{-2}$\\
         $k_3$ & 6.15 mg s${}^{-2}$& 19.6 mg s${}^{-2}$\\
         \hline
    \end{tabular}\]
    \caption{Parameter values used to generate Fig.~2 of the main text.}
    \label{tab:parameter_vals}
\end{table}

For Fig.~2(a)-(d), we use parameter values consistent with a spherical polystyrene bead of 1 micron diameter. The value of $\zeta$ is chosen to be consistent with the bead suspended in water at 300K and undergoing Stokes flow. We note that viscosity itself is a temperature dependent quantity, though we leave the consideration of these small corrections to future work. For Fig.~2(e)-(h) in contrast, we consider a denser and larger bead, such that the thermodynamic costs of the resulting protocols are more easily distinguished.

\section{Numerical validation of slow-driving approximation}
In this section, we expand on the definition of the slow-driving approximation and compare the results of this approximation to the numerical results from simulation. In essence, the crux of this regime is that all driving timescales, given by $\bm{\tau}_{D,i} \sim |\Lambda_i/\dot{\Lambda}_i|$, occur slower than all relaxation timescales for the system in question. Under this assumption and following standard linear response theory \cite{2012_PRL_Sivak} or alternatively a derivative truncation approximation \cite{2012_PRE_zulkowski} leads to the definition of the thermodynamic metric tensor given by Eq.~(5) of the main text and Eq~(S1) of the supplement.

For the model system in question, there are two relaxation timescales relevant to driving: the inertial, underdamped timescale $\tau_u = m/\zeta$ and the overdamped timescale $\tau_o = \zeta/k$. For driving that occurs significantly slower than this timescale, the assumptions underpinning thermodynamic geometry become more accurate and so too should its predictions. However, $\tau_o$ itself depends on the current value of the control parameter $k$, as does the characteristic maximum driving timescale $\tau_D$, set by the maximum value of $\{|\Lambda_i/\dot{\Lambda}_i|\}$. Therefore, for protocols that occur over a variety of values of $k$ and $T$, thermodynamic geometric increasingly becomes accurate for $\tau_u/\tau_D, \tau_o/\tau_D \ll 1$ at all times over the full protocol. This is especially true of steady-state engines for which accumulated distances from equilibrium compile over time. To this end, we define the dimensionless quantities 
\begin{equation}
A = \frac{1}{\tau} \oint_0^\tau \frac{\tau_u}{\tau_D(t)} dt,  \ \ \ B =  \frac{1}{\tau}\oint_0^\tau \frac{\tau_o(t)}{\tau_D(t)} dt
\end{equation}
where, as in the main text, $\tau$ represents the protocol duration of a full cycle, and $A$ and $B$ serve as time averages of the ratios of the time scales. As  before, thermodynamic geometry will prove more accurate for $A,B\ll 1$. Finally, defining the fractional time $s\equiv t/\tau$ and the non-dimensionalized driving timescale $\tilde{\tau}_D(s) \equiv \tau_D(s\tau)/\tau$, we have
\[A = \frac{1}{\tau} \oint_0^\tau \frac{\tau_u}{\tau_D(t)} dt =\frac{ \oint_0^1 \frac{\tau_u}{\tilde{\tau}_D(s)}}{\tau} \equiv \frac{\tau_A}{\tau}, \ \ \ B =  \frac{1}{\tau}\oint_0^\tau \frac{\tau_o(t)}{\tau_D(t)} dt = \frac{ \oint_0^1 \frac{\tau_o(s\tau)}{\tilde{\tau}_D(s)}}{\tau} \equiv \frac{\tau_B}{\tau},\]
where we define 
\begin{equation}
\tau_A \equiv  \oint_0^1 \frac{\tau_u}{\tilde{\tau}_D(s)}, \ \ \ \tau_B \equiv \oint_0^1 \frac{\tau_o(s\tau)}{\tilde{\tau}_D(s)}
\end{equation}
both of which are timescales independent of $\tau$ such that for $\tau_A,\tau_B \ll \tau$, thermodynamic geometry is a good approximation. We can now calculate $\tau_A$ and $\tau_B$ for all protocols considered, which we show in Table~S2:

\begin{table}[h!]
    \centering
\begin{tabular}{| c || c | c | c | c |}
\hline
\multirow{2}{4em}{Protocol} &  \multicolumn{2}{|c|}{Parameters of Fig.~2(a)-(d)}&   \multicolumn{2}{|c|}{Parameters of Fig.~2(e)-(h)} \\
\cline{2-5}
& $ \tau_A $ (s) & $\tau_B$ (s) &$ \tau_A$ (s) &$ \tau_B$ (s)\\ 
\hline\hline
Optimal Carnot & $3.33755\times10^{-7}$&0.00676226 & \ \ \ \ \  24.8006 \ \ \ \ \  & 4.22641 \\
\hline
Optimal geodesic &$3.33755\times10^{-7}$ & 0.00676226 & \ \ \ \ \ 33.6818\ \ \ \ \  & 6.52093\\
\hline
Hybrid &$3.33755\times10^{-7}$ &0.00676226& \ \ \ \ \  26.6871  \ \ \ \ \ & 5.12534\\
\hline
Protocol of \cite{2016_NatPhys_Martinez} & $3.33755\times10^{-7}$&0.00676226 & \ \ \ \ \ 24.8006 \ \ \ \ \ & 4.22641\\
\hline
\end{tabular}
\caption{Timescales $\tau_A$ and $\tau_B$ at which slow-driving and thermodynamic geometry becomes a valid approximation. The larger of $\tau_{A/B}$ is shown in dotted lines in Fig.~\ref{fig:LR}}
\label{fig:timescales}
\end{table}

\begin{figure}[h]
\centering{
\includegraphics[width=\columnwidth]{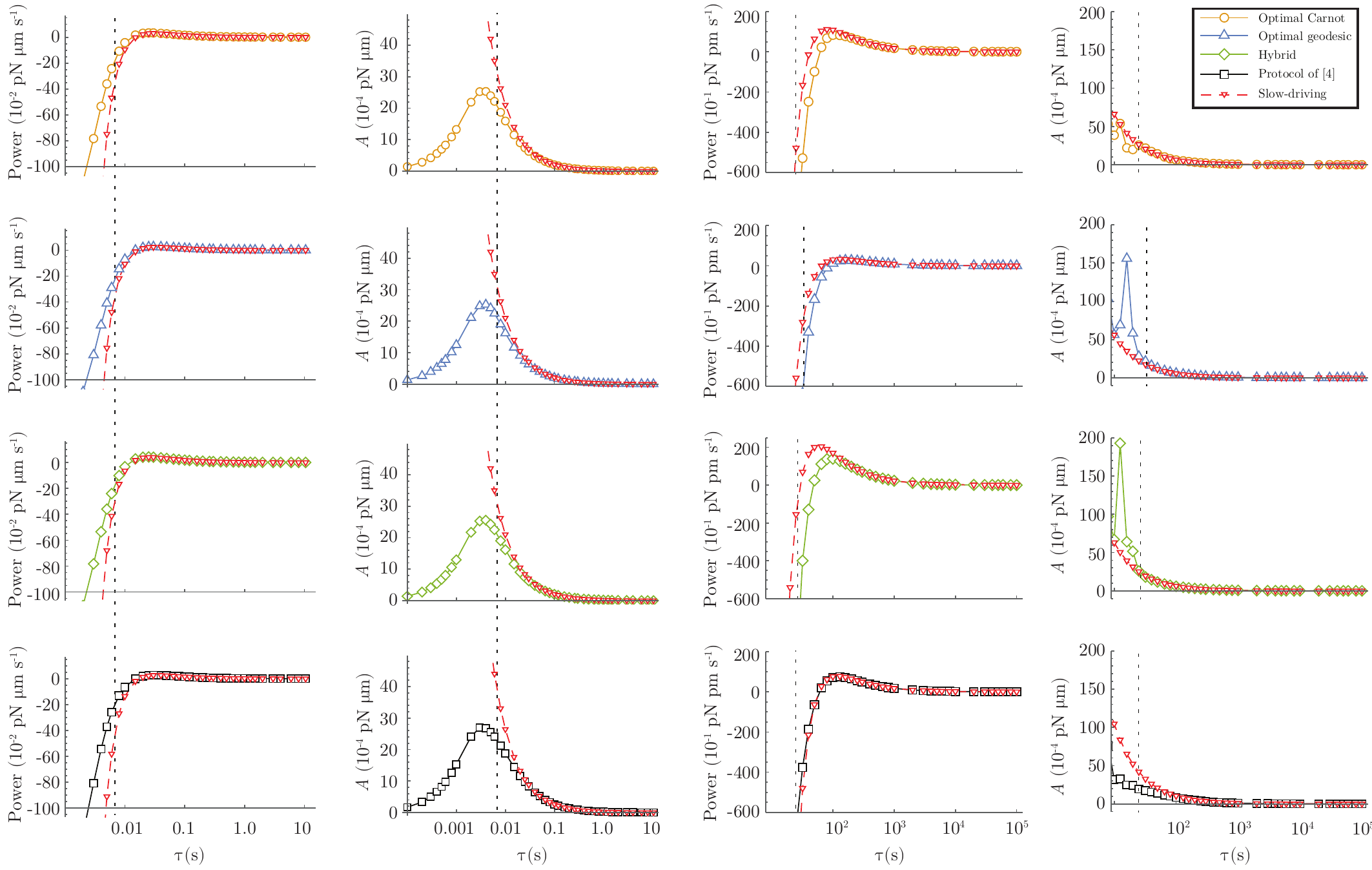}
 \\}
 \caption{Comparison of full numerical results to linear response: Power and dissipated availibility $A$ as a function of protocol duration for optimal Carnot (orange circles), geodesic (blue triangles), hybrid (green diamonds), and simulations of the experimental protocol used in \cite{2016_NatPhys_Martinez} (black squares), compared against the calculated linear response result for each (red down triangles). The linear response result is calculated for each protocol specifically. Dashed lines represent the larger of $\tau_{A/B}$ and therefore the timescale at which linear response is expected to start to be predictive. The left two columns are for overdamped parameters, same as Fig.~2(a)-(d) of main text, and the right two columns are for underdamped parameters, same as Fig.~2(e)-(h) of main text, see Table~\ref{tab:parameter_vals} for specific material values. All subplots in the same column share  horizontal axes.
  }
\label{fig:LR}
\end{figure}

With these results in hand, we now plot the numerical simulation results of dissipated energy and power as a function of protocol duration on the same axes as the linear response result (in red upside down triangles). Each subplot contains the simulation results for a single quantity for a single protocol (e.g. the Hybrid protocol is in green diamonds on the third row) and is plotted against the linear response prediction for the corresponding protocol: importantly, although all linear response predictions are shown with identical red triangles, the linear response prediction in each subplot is the one specifically matched to the numerical protocol shown in the same subplot (i.e. they are \textit{not} all the same). The two left columns correspond to similar material parameters as \cite{2016_NatPhys_Martinez} and Fig.~2(a)-(d) of the main text and the two right columns correspond to the same as Fig.~2(e)-(h) of the main text. The vertical dotted lines are the maximum of $\tau_A,\tau_B$ for each plot. For $\tau \gg \tau_A,\tau_B$ indeed we see excellent agreement with the linear response predictions and often even relatively good agreement for $\tau \gtrsim \tau_A,\tau_B$, showing the significant predictive power for this theory over a wide range of protocol durations and further justifying its popularity as one perturbative approach to finite-time thermodynamics.

\section{Comparison of simulated results to experiment}
\begin{figure}[h]
\centering{
\includegraphics[width=\columnwidth]{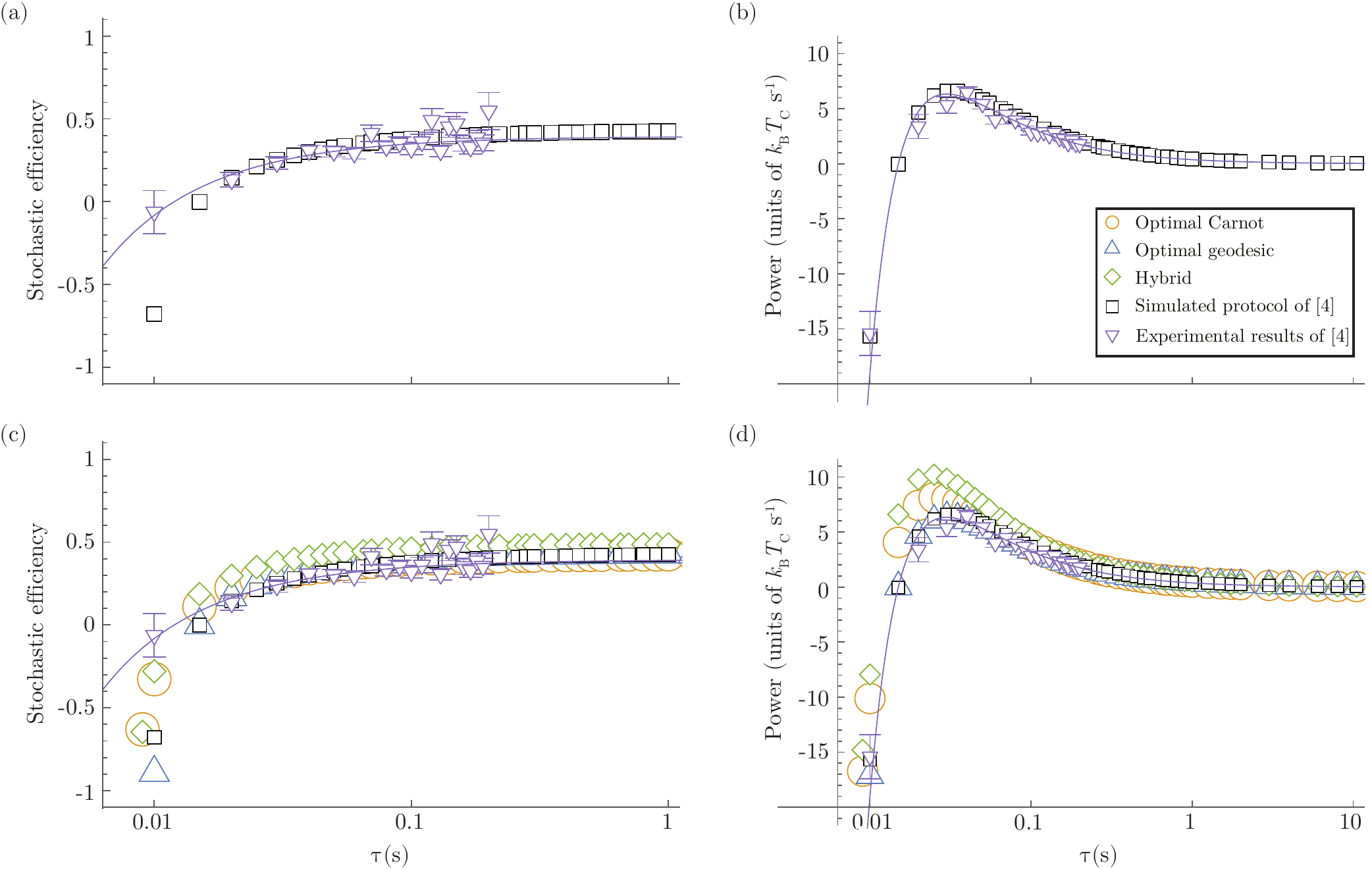}
 \\}
 \caption{\textbf{(a)} Stochastic efficiency and \textbf{(b)} power as a function of protocol duration for simulations of the experiment in \cite{2016_NatPhys_Martinez} (black squares) compared against experimental data provided from the authors of \cite{2016_NatPhys_Martinez} (purple down triangles). Error bars are the standard error of this data and the purple line is a best-fit (cf. Fig. 2b of \cite{2016_NatPhys_Martinez}). \textbf{(c)} and \textbf{(d)} same as \textbf{(a)} and \textbf{(b)}, respectively, but now with optimal Carnot (orange circles), geodesic (blue triangles), hybrid (green diamonds) for same material parameters also shown. Horizontal axes of \text{(a)} and \text{(b)} are the same as \text{(c)} and \textbf{(d)}, respectively.
 }
\label{fig:exp}
\end{figure}

In the main text, we showed the results of simulations for cycles operating in steady state for the colloidal parametric harmonic oscillator system studied previously in experiment \cite{2016_NatPhys_Martinez,2006_PRL_Blickle}, and used simulations of the Carnot cycle protocol studied in \cite{2016_NatPhys_Martinez} as a benchmark. To validate these simulations, here we compare directly with the experiment itself. In Fig.~\ref{fig:exp}(a)-(b), we display the stochastic efficiency and power as a function of protocol duration for experimental data provided by the authors of \cite{2016_NatPhys_Martinez} plotted against the numerical simulation meant to replicate this experiment. We find good agreement throughout the experimental regime tested but now with further quantitative predictions beyond this regime. In addition, in Fig.~\ref{fig:exp}(c)-(d), we plot the same thermodynamic quantities but now for the experimental results plotted against all the all numerically simulated cycles again using similar material parameters to those used experimentally. We now see that the optimized Carnot and hybrid engines outperform not only the averages of the experimental results but also the uncertainties for a large range of protocol durations, further bolstering our results. 

It bears mentioning that the stochastic efficiency is a further different definition of efficiency, introduced in \cite{2016_NatPhys_Martinez}. To reiterate, the efficiency we consider in the main text, which we will term ``irreversible efficiency" to avoid confusion, is defined as $\eta \equiv W/U \leq 1$ where $W$ is output work and $U$ is input thermal energy, which has an upper bound of unity saturated for \textit{any} reversible cycle. The standard ``textbook" definition of efficiency, which we term ``thermodynamic efficiency" $\eta_\text{th}$ is given by the ratio of $W/Q_+$ where $Q_+$ is the rectified heat flow into the system over the full cycle (i.e. only heat flow in is counted). For the quasistatic Carnot cycle, the thermodynamic efficiency is by definition $W/Q_H$ where $Q_H$ is the heat flow into the system from the hot reservoir as all other strokes are either adiabatic, $Q=0$, or have a heat flow purely out of the system. However, for finite-time Carnot engines, this is not necessarily true, such that \cite{2016_NatPhys_Martinez} defines the ``stochastic efficiency" as $\eta_s = W/(Q_1^{\text{adia}}+Q_H+Q_2^{\text{adia}})$, where $Q_{1/2}^{\text{adia}}$ are total (non-rectified) heat flow during the adiabatic strokes. They show that this definition of efficiency converges faster to the quasistatic value of $\eta_\text{th}$ as a function of protocol duration and so study it in their experiment. This quantity is what is plotted in Fig.~\ref{fig:exp}(a),(c).